\documentclass[a4paper,fleqn,usenatbib]{mnras}
\usepackage[T1]{fontenc}
\usepackage{ae,aecompl}
\usepackage{graphicx}
\usepackage{amsmath}
\usepackage{amssymb}
\usepackage{txfonts}

\newcommand{\ltsima}{$\; \buildrel < \over \sim \;$}
\newcommand{\lsim}{\lower.5ex\hbox{\ltsima}}
\newcommand{\gtsima}{$\; \buildrel > \over \sim \;$}
\newcommand{\gsim}{\lower.5ex\hbox{\gtsima}}
\newcommand{\bra}{\langle}
\newcommand{\ket}{\rangle}

\newcommand{\dd}{\mathrm{d}}

\newcommand{\Om}{\Omega_\mathrm{m}}
\newcommand{\de}{\mathrm{d}}

\newcommand{\ci}{\mathrm{i}}

\newcommand{\trace}{\mathrm{tr}}

\graphicspath{{figures/}}

\title[Separation of gravitational lensing and IAs]
{Statistical separation of weak gravitational lensing and intrinsic ellipticities based on galaxy colour information}
\author[T.M. Tugendhat, R. Reischke, B.M. Sch{\"a}fer]
{Tim M. Tugendhat, Robert Reischke, Bj{\"o}rn Malte Sch{\"a}fer\thanks{e-mail: bjoern.malte.schaefer@uni-heidelberg.de}\\
Astronomisches Rechen-Institut, Zentrum f{\"u}r Astronomie der Universit{\"a}t Heidelberg, Philosophenweg 12, 69120 Heidelberg, Germany}

\begin{document}
\pagerange{\pageref{firstpage}--\pageref{lastpage}}
\pubyear{2018}
\maketitle
\label{firstpage}

\begin{abstract}
Intrinsic alignments of galaxies are recognised as one of the most important systematic in weak lensing surveys on small angular scales. In this paper we investigate ellipticity correlation functions that are measured separately on elliptical and spiral galaxies, for which we assume the generic alignment mechanisms based on tidal shearing and tidal torquing, respectively. Including morphological information allows to find linear combinations of measured ellipticity correlation functions which suppress the gravitational lensing signal completely or which show a strongly boosted gravitational lensing signal relative to intrinsic alignments. Specifically, we find that $(i)$ intrinsic alignment spectra can be measured in a model-independent way at a significance of $\Sigma\simeq 60$ with a wide-angle tomographic survey such as Euclid's, $(ii)$ the underlying intrinsic alignment model parameters can be determined at percent-level precision, $(iii)$ this measurement is not impeded by misclassifying galaxies and assuming a wrong alignment model, $(iv)$ parameter estimation from a cleaned weak lensing spectrum is possible with almost no bias and $(v)$ the misclassification would not strongly impact parameter estimation from the boosted weak lensing spectrum.
\end{abstract}


\section{Introduction}
The next generation of weak lensing surveys has the goal to measure correlations in the shapes of neighbouring galaxies over a wide range of angular scales and with resolution in redshift in order to investigate the properties of gravity through its influence on cosmic structure formation and in order to carry out a precision determination of cosmological parameters, investigate the properties of gravity on large scales or determine details of dark energy models through their influence on cosmic structure formation. While weak lensing surveys like Euclid or LSST provide exquisite statistical precision due to the vast amount of data, the control of systematics is the primary obstacle in the way of exploiting this information. 

Weak cosmic shear refers to the correlated distortion of the cross-section of light bundles that reach us from distant galaxies \citep{mellier_probing_1999, bartelmann_weak_2001, refregier_weak_2003, hoekstra_weak_2008, munshi_cosmology_2008, kilbinger_cosmology_2015}. Using it as a cosmological probe one primarily determines angular correlations in the ellipticities of galaxies, after subdividing the galaxy sample in redshift bins. Under the assumption of intrinsically uncorrelated shapes the angular ellipticity correlation function measures lensing-efficiency weighted tidal shear correlations in the cosmic large-scale structure. These correlations carry information about the structure formation history as well as the expansion history of the Universe \citep{matilla_geometry_2017}, and do not rely on any assumption apart from the validity of a gravitational theory \citep{huterer_weak_2002, Linder2003, amendola_measuring_2008, bernstein_comprehensive_2009, huterer_weak_2010, carron_probe_2011, martinelli_constraining_2011, vanderveld_testing_2012}.

Intrinsic alignments of galaxies mimic correlations in the shapes of neighbouring galaxies which would be naively contributed to gravitational lensing \citep[for reviews, ][]{2015arXiv150405465K, 2015arXiv150405546K, 2015SSRv..tmp...65J, schaefer_review:_2009, troxel_intrinsic_2015}. If unaccounted for, either by modelling or by mitigation, they would interfere with the parameter inference process and would lead to wrong conclusions about the cosmological model \citep{kirk_impact_2010, kirk_optimising_2011, kirk_cosmological_2012, laszlo_disentangling_2012, capranico_intrinsic_2013, schaefer_angular_2015, krause_impact_2016, tugendhat_angular_2018}. While the exact mechanisms of galaxy alignment with the cosmic large-scale structure are not yet clear, tidal alignment models provide a physically motivated way to link the shapes of galaxies to the matter distribution on large-scales. In the case of elliptical galaxies, for which the linear alignment model (also called tidal alignment model) might be applicable, one assumes a distortion of the galaxy ellipsoid by tidal gravitational forces, which act perturbatively on the galaxy and exert a shearing distortion \citep{hirata_intrinsic_2004, hirata_intrinsic_2010, blazek_testing_2011, blazek_beyond_2017}. In this case, the observed ellipticity is proportional to the tidal shear components perpendicular to the line of sight. In contrast, the alignment of spiral galaxies may be due to the quadratic alignment model, where the orientation of the galaxy is linked to the host halo angular momentum, which in turn is generated in the early stages of structure formation by tidal torquing \citep{crittenden_spin-induced_2001, natarajan_angular_2001, mackey_theoretical_2002}. As an orientation effect, the alignment of spiral galaxies would not reflect the magnitude of tidal fields but would only depend on their orientation. Applying the models in the strictest sense would result in predicting a non-vanishing cross-correlation between the gravitational lensing effect and the linear tidal alignment of elliptical galaxies.

Our study is motivated by the fact that the distortion of galaxy ellipticities due to gravitational lensing should be universal and not depend on galaxy type. In this case one should be able to make use of morphological information in order to find linear combinations of ellipticity maps where the intrinsic alignment signal is up- or down-weighted relative to the gravitational lensing signal. We will investigate the usability of ellipticity spectra with weighted relative contributions from gravitational lensing and from intrinsic alignments, with the purpose of cosmological inference from weak lensing with suppressed IA-induced biases as well as for investigating alignment signals with a suppressed weak gravitational lensing effect, that would otherwise dominate ellipticity correlations.

The fiducial cosmological model is a spatially flat $\Lambda$CDM-cosmology motivated by the Planck-results \citep{2015arXiv150201590P}, with specific parameter choices $\Omega_m = 0.32$, $n_s = 0.96$, $\sigma_8 = 0.83$ and $h=0.68$, with a constant dark energy equation of state parameter of $w=-1.0$. We adopt the summation convention and orient coordinate systems in such a way that the $z$-axis points along the line of sight. We carry out all computations and statistical estimates for the Euclid-missions, but all statements would be applicable in a similar way to other weak lensing surveys with similar survey depths. After a summary of cosmology in Sect.~\ref{sect_cosmology} we review weak gravitational lensing and intrinsic alignments in Sect.~\ref{sect_lensing}. We introduce our method and demonstrate the separation technique in Sect.~\ref{sect_separation} and summarise and discuss our results in Sect.~\ref{sect_summary}.

\section{cosmology}\label{sect_cosmology}
Under the symmetry assumption of Friedmann-Lema{\^i}tre-cosmologies all fluids are characterised by their density and their equation of state: In spatially flat cosmologies with the matter density parameter $\Om$ and the corresponding dark energy density $1-\Om$ one obtains for the Hubble function $H(a)=\dot{a}/a$ the expression,
\begin{equation}
\frac{H^2(a)}{H_0^2} = \frac{\Om}{a^{3}} + \frac{1-\Om}{a^{3(1+w)}},
\end{equation}
The comoving distance $\chi$ is related to the scale factor $a$ through
\begin{equation}
\chi = -c\int_1^a\:\frac{\dd a}{a^2 H(a)},
\end{equation}
where the Hubble distance $\chi_H=c/H_0$ sets the distance scale for cosmological distance measures. Small fluctuations $\delta$ in the distribution of dark matter grow, as long as they are in the linear regime $\left|\delta\right|\ll 1$, according to the growth function $D_+(a)$ \citep{Linder2003},
\begin{equation}
\frac{\dd^2}{\dd a^2}D_+(a) +
\frac{2-q}{a}\frac{\dd}{\dd a}D_+(a) -
\frac{3}{2a^2}\Om(a) D_+(a) = 0,
\label{eqn_growth}
\end{equation}
and their statistics is characterised by the spectrum $\bra \delta(\bmath{k})\delta^*(\bmath{k}^\prime)\ket = (2\pi)^3\delta_D(\bmath{k}-\bmath{k}^\prime)P_\delta(k)$. Inflation generates a spectrum of the form $P_\delta(k)\propto k^{n_s}T^2(k)$ with the transfer function $T(k)$ \citep{1986ApJ...304...15B} which is normalised to the variance $\sigma_8$ smoothed to the scale of $8~\mathrm{Mpc}/h$,
\begin{equation}
\sigma_8^2 = \int_0^\infty\frac{k^2\dd k}{2\pi^2}\: W^2(8~\mathrm{Mpc}/h\times k)\:P_\delta(k),
\end{equation}
with a Fourier-transformed spherical top-hat $W(x) = 3j_1(x)/x$ as the filter function. From the CDM-spectrum of the density perturbation the spectrum of the dimensionless Newtonian gravitational potential $\Phi$ can be obtained,
\begin{equation}
P_\Phi(k) \propto \left(\frac{3\Om}{2 (k\chi_H)^2}\right)^2\:P_\delta(k)
\end{equation}
by applying the comoving Poisson-equation $\Delta\Phi = 3\Om/(2\chi_H^2)\delta$ for deriving the gravitational potential $\Phi$ (in units of $c^2$) from the density $\delta$. Because our analysis relies on the assumption of Gaussianity, we need to avoid nonlinearly evolving scales and will restrict our analysis to large angular and spatial scales, where the cosmic density field can be approximated to follow a linear evolution, conserving the near-Gaussianity of the initial conditions. We increase the variance of the weak lensing signal and of the intrinsic alignment signal of elliptical galaxies on small scales because of nonlinear structure formation using the description of \citet{Smith:2002dz, casarini_non-linear_2011, casarini_tomographic_2012}. Consequently, the cross-correlation between weak lensing and elliptical galaxy shapes will likewise have increased variances on small scales. Shapes of spiral galaxies are set by the initial conditions of structure formation, therefore we did not apply changes to the linear CDM-spectrum $P(k)$.

\section{weak gravitational lensing and intrinsic alignments}\label{sect_lensing}

\subsection{Gravitational tidal fields and their statistics}
In our investigation, alignments of galaxies with the large-scale structure are due to gravitational-tidal interactions with shear fields. These fields will be assumed to have Gaussian statistics. This assumption will not be applicable at late times and on small scales. Tidal alignment models relate correlations in the shapes of galaxies to correlations in the tidal shear field, 
\begin{equation}
\Phi_{\alpha\beta}(\bmath{x}) = \frac{\partial^2\Phi(\bmath{x})}{\partial x_\alpha\partial x_\beta},
\end{equation}
as a tensor containing the second derivatives of the Newtonian gravitational potential. Correlations of $\Phi_{\alpha\beta}(\bmath{x})$ as a function of distance $r=\left|\bmath{x}-\bmath{x}^\prime\right|$ will be described by the correlation function 
\begin{equation}
C_{\alpha\beta\gamma\delta}(r)\equiv\bra\Phi_{\alpha\beta}(\bmath{x})\Phi_{\gamma\delta}(\bmath{x}^\prime)\ket
\end{equation}
which \citet{catelan_correlations_2001} have shown to take the form
\begin{equation}
\begin{split}
C_{\alpha\beta\gamma\delta}(r)  = & \
(\delta_{\alpha\beta}\delta_{\gamma\delta}+\delta_{\alpha\gamma}\delta_{\beta\delta}+\delta_{\alpha\delta}\delta_{\beta,\gamma})\:\zeta_2(r) \\ 
& + (\hat{r}_\alpha \hat{r}_\beta \delta_{\gamma\delta}+\mathrm{5~perm.})\zeta_3(r)+
\hat{r}_\alpha \hat{r}_\beta \hat{r}_\gamma \hat{r}_\delta\:\zeta_4(r).
\label{eqn_decomp}
\end{split}
\end{equation}
The fluctuation statistics of the gravitational potential $P_\Phi(k)$ enters through the functions $\zeta_n(r)$,
\label{sec:quadalign}
\begin{equation}
\zeta_n(r) = \left(-1\right)^n r^{n-4}\int\frac{\de{k}}{2\pi^2}\:P_\Phi(k)\,k^{n+2}\,j_n(kr),
\label{eq:zeta_n}
\end{equation}
as derived by \citet{crittenden_spin-induced_2001}. $\hat{r}_\alpha$ is the $\alpha$-component of the unit vector parallel to $r=\bmath{x}-\bmath{x}^\prime$.

While the tidal shear fields will directly change the shape of an elliptical galaxy, the effect of tidal fields on a spiral galaxies is to determine its angular momentum direction and consequently the inclination of the galactic disc. In both cases, the relevant components of the tidal shear tensor are those of the traceless part. The resulting correlation function  $\tilde{C}_{\alpha\beta\gamma\delta}(r)$ of the traceless tidal shear $\tilde{\Phi}_{\alpha\beta} = \Phi_{\alpha\beta} - \Delta\Phi/3\:\times\delta_{\alpha\beta}$, would be given by
\begin{equation}
\begin{split}
\tilde{C}_{\alpha\beta\gamma\delta}(r)&  =  \ 
C_{\alpha\beta\gamma\delta}(r) \\
& - \frac{1}{3}\left(\delta_{\gamma\delta}\left(5\zeta_2(r)+\zeta_3(r)\right) + \hat{r}_\gamma \hat{r}_\delta\left(7\zeta_3(r)+\zeta_4(r)\right)\right)\delta_{\alpha\beta} \\
& - \frac{1}{3}\left(\delta_{\alpha\beta}\left(5\zeta_2(r)+\zeta_3(r)\right) + \hat{r}_\alpha \hat{r}_\beta\left(7\zeta_3(r)+\zeta_4(r)\right)\right)\delta_{\gamma\delta} \\
& + \frac{1}{9}\left(15\zeta_2(r)+10\zeta_3(r)+\zeta_4(r)\right)\delta_{\alpha\beta}\delta_{\gamma\delta}.
\end{split}
\end{equation}

Furthermore, the apparent shapes of spiral galaxies only depend on the direction of the angular momentum and not its magnitude because their ellipticity is only an orientation effect. Consequently, their shape correlations can be traced back to the traceless, unit-normalised tidal shear field $\hat{\Phi}_{ij}$, which obeys the conditions $\hat{\Phi}_{ii}=0$ and $\hat{\Phi}_{ij}\hat{\Phi}_{ji}=1$. It can be decomposed into correlations $\tilde{C}_{AB} = \bra\tilde{\Phi}_A\tilde{\Phi}_B\ket$ of the traceless tidal shear $\tilde{\Phi}_A$ field \citep{natarajan_angular_2001, crittenden_spin-induced_2001} by virtue of Wick's theorem,
\begin{equation}\label{eq:fourpoint}
\bra\hat{\Phi}_A(\bmath{x})\hat{\Phi}_B(\bmath{x})\:\hat{\Phi}_C(\bmath{x}^\prime)\hat{\Phi}^\prime_D(\bmath{x}^\prime)\ket =
\frac{1}{\left[14\zeta_2(0)\right]^2}\left(\tilde{C}_{AC}\tilde{C}_{BD}+\tilde{C}_{AD}\tilde{C}_{BC}\right),
\end{equation}
where $A$, $B$, $C$ and $D$ are containers for pairs of indices. The normalisation with $\zeta_2$ has the consequence that correlations of the unit-normalised tidal shear field are only due to the orientation of the eigen-systems and do not depend on the absolute magnitude of the tidal shear. The assumption of Gaussianity of the aligning large-scale structure is in fact a strong one, as alignments on filaments has been demonstrated with numerical simulations \citep{codis_intrinsic_2015}. To what extend these alignments would, after Limber-projection, differ from those derived from a Gaussian random field, has not yet estimated. Likewise, depending on the details of halo-model based intrinsic alignment models \citep{joachimi_intrinsic_2013,joachimi_intrinsic_2013-1}, a separation of spiral and elliptical alignment due to their dependence with powers of the tidal shear field would not be feasible.

\subsection{Weak gravitational lensing}
The lensing potential $\psi$ \citep[for a review see e.g.][]{bartelmann_weak_2001} is a line-of-sight projection of the gravitational potential $\Phi$:
\begin{equation}
\psi = \int_0^{\chi_\mathrm{H}}\mathrm{d}\chi W(\chi)\Phi,
\end{equation}
it thus inherits the statistical properties of the gravitational potential. The changes in size and shape of a light bundle are cause by differential deflection and are therefore proportional to the second derivatives of $\psi$ perpendicular to the line sight and can be decomposed in terms of the weak lensing convergence $\kappa$,
\begin{equation}
\kappa = 
\int\dd\chi\:W(\chi)\:\sigma^{(0)}_{\alpha\beta}\:\frac{\partial^2\Phi}{\partial x_\alpha\partial x_\beta},
\end{equation}
which depends on the trace of the tidal shear, and the complex weak lensing shear $\gamma$,
\begin{equation}
\gamma = \gamma_+ +\ci\gamma_\times = 
\int\dd\chi\:W(\chi)\left(\sigma^{(1)}_{\alpha\beta}+\ci\sigma^{(3)}_{\alpha\beta}\right)\frac{\partial^2\Phi}{\partial x_\alpha\partial x_\beta},
\end{equation}
which reflects the traceless part of the tidal shear. $\sigma_{\alpha\beta}^{(n)}$ are the Pauli-matrices, and effectively only the derivatives of $\Phi$ perpendicular to the line of sight are relevant. The weight function $W(\chi)$ is given by
\begin{equation}
W(\chi) = 2 \frac{D_+(\chi)}{a}G(\chi)\chi,
\end{equation}
with the lensing efficiency function
\begin{equation}
G(\chi) = \int_{\chi}\mathrm{d}\chi\:n(\chi')\frac{\mathrm{d}z}{\mathrm{d}\chi'}\left(1-\frac{\chi}{\chi'}\right). 
\end{equation}
with $\mathrm{d}z/\mathrm{d}\chi' = H(\chi')/c$ and $n(\chi')$ being the  distribution of the sources. Because weak gravitational lensing affects both elliptical and spiral galaxies alike, we scale the resulting lensing spectra $C^{\gamma}_{ij}(\ell)$ with the total number of galaxy pairs $\simeq n^2=(n_s+n_e)^2$. In our analysis, we are using a fit for the nonlinear power spectrum \citep{Smith:2002dz} to calculate the gravitational lensing contribution.

\subsection{Alignments of spiral galaxies}
In the tidal torque model used in this work, the alignment of spiral galaxies is purely due to their orientation, which in turn is related to the angular momentum correlation of neighbouring galaxies relative to the line of sight \citep{croft_weak-lensing_2000,crittenden_spin-induced_2001}. Angular momentum correlations are mainly build up at early times during structure formation and are thus due to initial correlations \citep{catelan_evolution_1996, theuns_angular_1997, catelan_non-linear_1997}. The correlated angular momenta result into to correlated inclination angles of neighbouring galaxies and thus ultimately into correlated ellipticities \citep{catelan_intrinsic_2001}. Assuming that the symmetry axis of the galactic disc coincides with the direction of the angular momentum $\hat{L} = \boldsymbol{L}/L$, the ellipticity can be written as
The alignment of spiral galaxies is purely due to the orientation of their circular disks, which in turn is related to the angular momentum correlation of neighbouring galaxies relative to the line of sight \citep{croft_weak-lensing_2000,crittenden_spin-induced_2001}. Angular momentum correlations are mainly build up at early times during structure formation and are thus due to initial correlations \citep{catelan_evolution_1996, theuns_angular_1997, catelan_non-linear_1997}. The correlated angular momenta result into to correlated inclination angles of neighbouring galaxies and thus ultimately into correlated ellipticities \citep{catelan_intrinsic_2001}. Assuming that the symmetry axis of the galactic disc coincides with the direction of the angular momentum $\hat{L} = \boldsymbol{L}/L$, the ellipticity can be written as
\begin{equation}\label{eq:ellipticity_angular}
\epsilon = \frac{\hat{L}^2_x - \hat{L}^2_y}{1+ \hat{L}^2_z} + 2\mathrm{i}\frac{\hat{L}_x\hat{L}_y}{1+\hat{L}^2_z}\; .
\end{equation} 
Angular momentum is generated by a torque exerted by the ambient large-scale structure onto the protogalactic halo, a mechanism called tidal torquing \citep{white_angular_1984,barnes_angular_1987,schaefer_review:_2009,stewart_angular_2013}. For Gaussian random fields the auto-correlation of angular momenta is given by \citep{lee_galaxy_2001}
\begin{equation}
\left\langle \hat{L}_\alpha \hat{L}_\beta \right\rangle =  \frac{1}{3}\left(\frac{1+ A}{3}\delta_{\alpha\beta} - A \hat\Phi_{\alpha\mu}\hat\Phi_{\mu\beta} \right)\; .
\end{equation}
The free parameter $A$ determines the strength of the coupling between alignment and tidal torque. Since the correlation is determined by the traceless part of the shear tensor $\hat\Phi_{\alpha\beta}$ the resulting effect is clearly due to orientation effects only. For a Gaussian distribution $p(\hat{L}|\hat{\Phi}_{\alpha\beta})\mathrm{d}\hat{L}$ and the use of eq.~(\ref{eq:ellipticity_angular}) one can express the ellipticity in terms of the tidal field
\begin{equation}
\epsilon(\hat\Phi) = 
\frac{A}{2}
\left(\hat{\Phi}_{x\alpha}\hat{\Phi}_{\alpha x} - \hat{\Phi}_{y\alpha}\hat{\Phi}_{\alpha y} -2\mathrm{i}\hat{\Phi}_{x\alpha}\hat{\Phi}_{\alpha y}\right)\; .
\end{equation}
Correlations in the ellipticities can thus be traced back to the 4-point function of the shear field, which is given in eq.~(\ref{eq:fourpoint}). For keeping a correct relative normalisation of the shape correlations, we scale the resulting angular ellipticity spectra $C^{\mathrm{s},II}_{ij}(\ell)$ with the squared number of spiral galaxies $n_s^2$. It is remarkable that the shapes of spiral galaxies in the quadratic alignment model are in fact sensitive to tidal shear components parallel to the line of sight, in fact, those components determine the magnitude of the alignment effect, in contrast to the alignment of elliptical galaxies in the linear alignment model or to gravitational lensing, which reflect purely the tidal shear components perpendicular to the line of sight. 

\begin{figure*}
\begin{center}
\includegraphics[width=1.7\columnwidth]{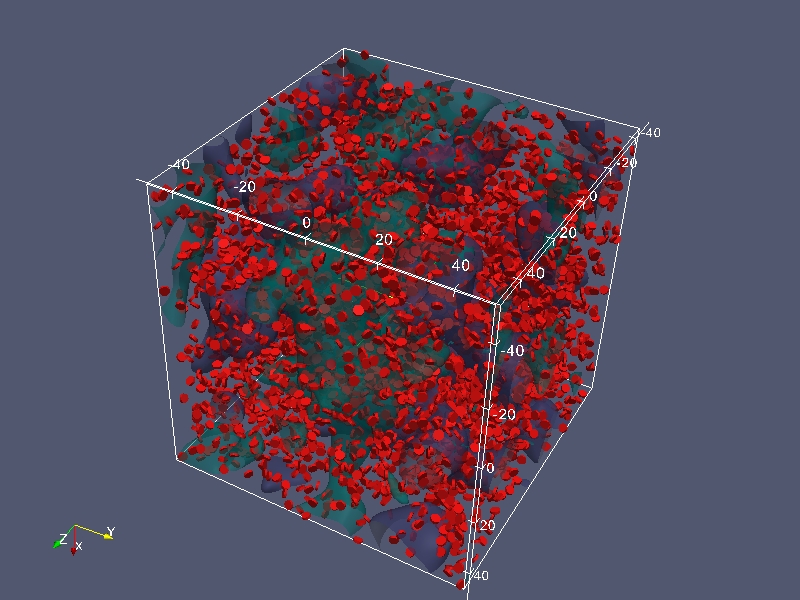}
\caption{A realisation of the cosmic matter distribution as a Gaussian random field in a cube of the side length $100~\mathrm{Mpc}/h$. The blue and green surfaces are $\pm1\sigma$-contours of the density field, and the orientation of the galactic discs (in red) follow from a quadratic, angular-momentum based alignment model.}
\label{fig_realisation}
\end{center}
\end{figure*}

\autoref{fig_realisation} gives a visual impression of alignments in the quadratic, angular momentum-based alignment model: From a realisation of a Gaussian random density field $\delta(\bmath{x})$ from the CDM-spectrum $P(k)$ we computed the traceless, unit-normalised tidal shear $\hat{\Phi}_{\alpha\beta}$, which is used to determine the variance of the distribution $p(\hat{L}_\alpha|\hat{\Phi}_{\alpha\beta})$. Angular momenta are drawn at random locations which fix the orientation of the galactic discs. The amount of correlation in the orientation of the galactic discs corresponds determined from the realisation corresponds to the theoretically computed correlation function. Ellipticity correlations between spiral galaxies are rather short-ranged, with a typical correlation length of about $1~\mathrm{Mpc}/h$ \citep{schafer_galactic_2012}, which makes them a small-scale phenomenon at angular scales of $\ell\simeq10^3$ for Gpc-scale surveys.

\subsection{Alignments of elliptical galaxies}
For elliptical galaxies we assume a virialised system in which stars move randomly in a gravitational potential $\Phi$ with the velocity dispersion $\sigma^2$. In equilibrium the density profile is a solution of the radially symmetric Jeans equation and scales $\rho\propto\exp(-\Phi/\sigma^2)$. In the presence of a tidal field induced by the ambient large-scale structure, the equilibrium situation is perturbed and the galaxy finds a new equilibrium. Perturbing the Jeans equation at first order in the tidal fields $\partial_\alpha\partial_\beta\Phi$ yields the following solution for the density:
\begin{equation}
\rho \propto \exp\left(-\frac{\Phi(\boldsymbol{x})}{c^2}\right)\left(1-\frac{1}{2\sigma^2}\frac{\partial^2\Phi(\boldsymbol{x})}{\partial x_\alpha\partial x_\beta}\bigg|_{\boldsymbol{x} = \boldsymbol{x_0}}x^\alpha x^\beta\right).
\end{equation}
While the reaction of the halo to the tidal fields is determined by the velocity dispersion $\sigma^2$, i.e. how strongly the particles are bound in the gravitational potential, the relationship between tidal shear field and ellipticity needs as well to reflect the luminous profile, which we absorb in the definition of a constant of proportionality $D$. Since this model gives rise to ellipticities being linear in the tidal fields it is commonly referred to as the linear alignment model \citep{hirata_intrinsic_2010, blazek_beyond_2017, blazek_separating_2012, blazek_testing_2011, blazek_tidal_2015}. Assuming $x$ and $y$ being coordinates perpendicular to the line of sight the complex ellipticity is given by
\begin{equation}
\epsilon = 
\epsilon_+ + \epsilon_\times = 
D\left(\frac{\partial^2\Phi}{\partial x^2}-\frac{\partial^2\Phi}{\partial y^2} + 2\mathrm{i}\frac{\partial^2\Phi}{\partial x\partial y}\right).
\end{equation}
From this equation it is clear that the linear alignment model depends on both the amplitude and the orientation of the tidal fields. In order to maintain the correct relative normalisation of the spectra, we scale the resulting $C^{\mathrm{e},II}_{ij}(\ell)$ with the number of elliptical galaxies $n_e^2$.

The alignment model parameters $A$ and $D$ are chosen to have values of $A = 0.25$ from numerical simulations of angular momentum generation in haloes, and $D = 9.5\times10^{-5} c^2$ from CFHTLenS-data, respectively. While $A$ is effectively a geometric, dimensionless parameter, $D$ links the dimensionless ellipticity to the tidal shear field. For a more detailed discussion of this relationship including the scaling with mass, we refer to \citet{piras_mass_2017} and \citet{tugendhat_angular_2018}. While there is no definitive measurement of ellipticity correlations of spiral galaxies, the case is notably different for elliptical galaxies \citep{mandelbaum_detection_2006, mandelbaum_wigglez_2011, 2015MNRAS.450.2195S}.

\subsection{Cross-alignments between intrinsic shapes and lensing}
For spiral galaxies there exist no GI-type terms due to Wick's theorem, because those terms would be proportional to a third moment of the tidal shear field. In contrast, there will be a non-vanishing cross-correlation between lensing and the intrinsic alignment of elliptical galaxies. We would like to point out that these cross-correlations, $C^{\mathrm{e},GI}_{ij}(\ell)$, have to be symmetrised with respect to the bin numbers: Naturally, the more distant galaxy is lensed whereas the closer galaxy is intrinsically aligned while the inverse is not possible. Statistical isotropy forces the covariance matrix to be symmetrical however \citep{tugendhat_angular_2018}. We scale the cross-spectra $C^{\mathrm{e},GI}_{ij}(\ell)$ with the number $n_e(n_e+n_s)=n_en$ of pairs involving at least one elliptical galaxy.

It is worth pointing out that there is a straightforward physical difference between the weak lensing and intrinsic alignments. As weak lensing is an integrated effect, there will be nonzero cross-correlations between different tomographic bins, whereas intrinsic alignments will, due to their locality, only show correlations within the same bin. This can already be used as a method of discrimination between $II$- and $GG$-spectra \citep{bernstein_dark_2004, huterer_nulling_2005}, but will not get rid of the $GI$-contribution.

\subsection{Shape correlations in a weak lensing survey}
We carry out our investigation for a weak lensing survey similar to Euclid's: The redshift distribution $n(z)\dd z$ is assumed to have the shape,
\begin{equation}
n(z)\dd z \propto \left(\frac{z}{z_0}\right)^2\exp\left[-\left(\frac{z}{z_0}\right)^\beta\right]\dd z,
\end{equation}
with the choices $\beta=3/2$ and $z = 1/\sqrt{2}$, which generates a median redshift of unity \citep{laureijs_euclid_2011}. In this work, we have used a yield of 40 galaxies per squared arcminute and to assume a fraction of $f_\mathrm{sky} = 0.5$ of the sky is observed, which is above the current expected specifications for Euclid.

Linking the correlations of the observable ellipticity $\epsilon$ to the tidal shear fields allows to express correlations in $\epsilon$ in terms of correlations in $\partial^2_{\alpha\beta}\Phi$. A suitable Limber-projection with the redshift-distribution $n(z)\dd z$ while introducing a binning allows us to compute angular correlation functions and in the next step, to obtain tomographic angular $E$-mode spectra $C^{\mathrm{s},II}_{ij}(\ell)$, $C^{\mathrm{e},II}_{ij}(\ell)$ and lastly $C^{\mathrm{e},GI}_{ij}(\ell)$, which we can compare to the tomographic weak lensing spectrum $C^\gamma_{ij}(\ell)$ \citep{hu_power_1999, hu_power_2001}. Due to the locality of intrinsic alignments, the two $II$-spectra are $\propto\delta_{ij}$, while the $GI$-spectrum or the weak lensing spectrum do not possess this property \citep{hirata_intrinsic_2004}. Central to our investigation will be the linear dependence of the ellipticity with tidal shear field for gravitational lensing and for the intrinsic alignment of elliptical galaxies, while the shapes of spiral galaxies depend on squares of the tidal shear. For non-tomographic, 3-dimensional weak lensing surveys, the effects of intrinsic alignments are physically identical and can be studied analogously \citep{merkel_intrinsic_2013}.

\begin{figure}
\begin{center}
\includegraphics[width=\columnwidth]{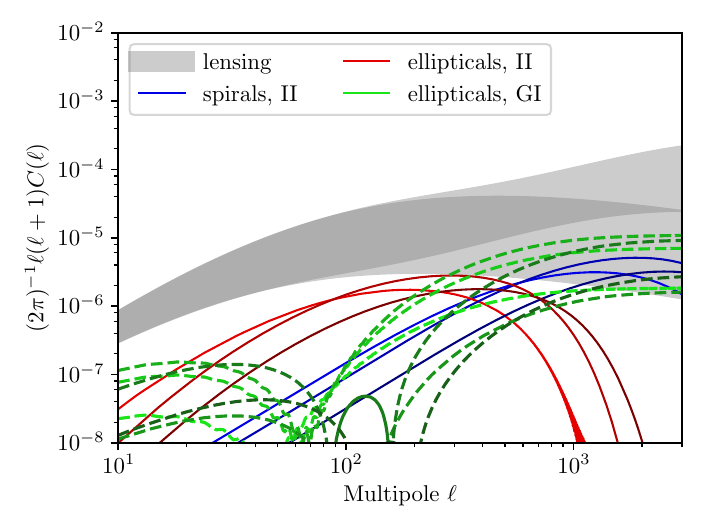}
\caption{Contributions to the ellipticity correlations by weak gravitational lensing $C^\gamma_{ij}(\ell)$. The number of tomography bins is $n_\mathrm{bin} = 3$, and negative contributions are depicted in dashed lines. The grey bands show the upper and lower limits of the six weak lensing spectra possible from $3$-bin tomography; separately for linear structure formation (lower band) and nonlinear structure formation (upper band), the linear lensing falls off at large $\ell$. The IA contributions are: $C^{\mathrm{e},GI}_{ij}(\ell)$ (green), $C^{\mathrm{s},II}_{ij}(\ell)$ (blue) and $C^{\mathrm{e},II}_{ij}(\ell)$ (red). Here, we also show all possible bin pairings, the three autocorrelations for the $II$-part, as well as all six possible combinations for the $GI$-part. Darker colour indicates higher bin numbers $ij$.}
\label{fig_spectra}
\end{center}
\end{figure}

\autoref{fig_spectra} shows the expected $E$-mode spectra for a $\Lambda$CDM-cosmology with a conventional choice of the alignment parameters $A$ and $D$ for tomography with $n_\mathrm{bin} = 3$ bins with Euclid, resulting from a Limber-projection \citep{limber_analysis_1954} and subsequent Fourier-transform. For a reasonably deep lensing survey such as Euclid's, lensing-induced ellipticity correlations dominate over intrinsic alignments. The IA contribution on large angular scales is caused by elliptical galaxies following the tidal shearing model, whereas on small scales the contribution from spiral galaxies, which is described by the tidal torquing model, is most important. Over a wide range of angular scales the negative cross-correlation between gravitational lensing and intrinsic alignments shapes the ellipticity spectrum. It is notable that the shapes of elliptical galaxies and lensing measure tidal field components perpendicular to the line of sight and are proportional to the magnitude of the tidal shears, but that in contrast spiral galaxies reflect with their shapes the tidal field orientation including line of sight-components. For details on the derivation of angular shape correlation functions and $E/B$-mode ellipticity spectra we refer to \citet{capranico_intrinsic_2013}, \citet{schaefer_angular_2015} and \citet{tugendhat_angular_2018}.

\section{Separation between weak lensing and galaxy alignments}\label{sect_separation}

\subsection{Idea and formalism}
Weak gravitational lensing and intrinsic alignments of galaxies are tidal gravitational effects of the cosmic large scale structure, but the details of how the observed shape of a galaxy is influenced by the tidal field $\partial^2_{ij}\Phi$ generated by the large-scale structure depends on the interaction mechanism: Gravitational lensing is universal and operates on all galaxy shapes in an identical way. For the intrinsic alignment contributions, we assume that elliptical galaxies change their shape in proportion to the tidal gravitational field according to the tidal shearing model and that the shape of spiral galaxies is an orientation effect that depends on the squared, unit-normalised tidal field, as stipulated by the tidal torquing model. The dependences on the linear and squared tidal shear field have the important consequence that there is no cross-correlation between lensing and the intrinsic shape of spiral galaxies, $\bra\gamma\epsilon_\mathrm{s}^\prime\ket = 0$ and neither between the shapes of spiral and elliptical galaxies, $\bra\epsilon_\mathrm{s}\epsilon_\mathrm{e}^\prime\ket = 0$, if the tidal shear field follows Gaussian statistics: In this case, the two correlation functions would be proportional to a third moment of a symmetric distribution, which makes them vanish. In contrast, there will be a nonzero cross-correlation between the intrinsic shapes of elliptical galaxies and weak lensing, $\bra\gamma\epsilon_\mathrm{e}^\prime\ket\neq 0$, which is usually dubbed GI-alignment. These model choices are consistent with \cite{tugendhat_angular_2018}, where the implications of such a physically motivated mixed model for the two overwhelmingly contributing morphologies is studied. Other model choices, such as the closely related \cite{blazek_beyond_2017}, wouldn't invalidate this method. Although correlations between the two morphologies would contaminate the isolated gravitational lensing signal, a separation of galaxy types using colour information would still be feasible. Using multiple models for each galaxy type would make the calculation more complex as well, but can be done in principle. We have chosen to use the aforementioned models for the two morphologies because we believe them to be a physically well-motivated choice and thus being the major contributor to the IA signal. A more complex model, for example additional linear term dependency of spirals at some scales, would amount to a misclassification of galaxy types in our ansatz, which we discuss in detail. Furthermore, an additional dependency of the spiral model on the amplitude of the tidal field would change the amplitude of the corresponding part of the IA signal, however it would not affect the points where both IA signals disappear. Superficially not unlike a multi-tracer method \citep[such as used in e.g.][]{Leonard_multiple_2018}, our formalism allows for a certain flexibility for which signals to boost or suppress respectively. For example, the lensing signal can be slightly enhanced over intrinsic alignments or eliminated completely depending only on the choice of a free parameter.

Starting with a tomographic observation of the ellipticity field in a range of redshift bins $i$, which contains contributions from weak gravitational lensing and from the two alignment mechanisms and which is sampled including colour-information we define the observed maps $e_{s,i}$ and $e_{e,i}$, where the subscripts denote spiral galaxies and elliptical galaxies. These contain contributions from weak lensing $\gamma_i$ in redshift bin $i$ and their respective alignment mechanism $\epsilon_{\mathrm{s},i}$ and $\epsilon_{\mathrm{e},i}$ and denoting the shape noise $\varepsilon$,
\begin{align}
e_{\mathrm{s},i} = \gamma_i + \epsilon_{\mathrm{s},i} + \varepsilon_{i},\\
e_{\mathrm{e},i} = \gamma_i + \epsilon_{\mathrm{e},i} + \varepsilon_{i},
\end{align}
respectively. Using the same value of $\gamma_i$ for both shapes assumes that the change in shape due to gravitational lensing does not depend on the type of galaxy. In reality, the situation is more complicated, because an estimate of the shape of the galaxy depends on the brightness distribution is a nonlinear process which is affected by the tidal shear and higher derivatives of the gravitational potential. Similarly, there are dependences of the measured shape on colour because of variations of the telescope's point spread function \citep{er_calibration_2018}.

Consequently, the covariance matrix $C_t(\ell) = S_t(\ell) + N_t(\ell)$ of a tomographic measurement of the data vector $(e_{\mathrm{s},i},e_{\mathrm{e},i^\prime})$,
\begin{equation}
C_t(\ell) =
\left(
\begin{array}{ll}
C^{\mathrm{ss}}_{ij}(\ell) & C^{\mathrm{se}}_{ij^\prime}(\ell)\\
C^{\mathrm{es}}_{i^\prime j}(\ell) & C^{\mathrm{ee}}_{i^\prime j^\prime}(\ell)
\end{array}
\right),
\end{equation}
is composed from the contributions of the signal $S_t(\ell)$
\begin{equation}
\begin{split}
&S_t(\ell) = \\
&\left(
\begin{array}{ll}
n_s^2\left(C^\gamma_{ij}(\ell) + C^{\mathrm{s},II}_{ij}(\ell)\right) &
n_sn_e\left(C^\gamma_{ij^\prime}(\ell) + C^{\mathrm{e},GI}_{ij^\prime}(\ell)\right) \\
n_sn_e\left(C^\gamma_{i^\prime j}(\ell) + C^{\mathrm{e},GI}_{i^\prime j}(\ell)\right) &
n_e^2\left(C^\gamma_{i^\prime j^\prime}(\ell) + 2C^{\mathrm{e},GI}_{i^\prime j^\prime}(\ell) + C^{\mathrm{e},II}_{i^\prime j^\prime}(\ell)\right)
\end{array}
\right),
\end{split}
\label{eqn_st}
\end{equation}
where the two intrinsic alignment autocorrelations $C^{\mathrm{s},II}_{ij}(\ell)$ and $C^{\mathrm{e},II}_{ij}(\ell)$ are diagonal and therefore proportional to $\delta_{ij}$ due to the locality of the tidal interaction process, and the noise $N_t(\ell)$,
\begin{equation}
N_t(\ell) = \sigma_\epsilon^2\:n_\mathrm{bin}\:
\left(
\begin{array}{ll}
n_\mathrm{s}\delta_{ij} &
0 \\
0 &
n_\mathrm{e}\delta_{i^\prime j^\prime}
\end{array}
\right),
\label{eqn_nt}
\end{equation}
which depends on the number $n_\mathrm{s}$ and $n_\mathrm{e}$ of spiral and elliptical galaxies, respectively. The dispersion $\sigma_\epsilon^2$ of the shape measurement is taken to be $\sigma_\epsilon= 0.3$ as expected from Euclid, and we set the number of bins to $n_\mathrm{bin} = 7$. The off-diagonal elements of $N_t(\ell)$ are zero because a galaxy can not be spiral and elliptical at the same time: If the ellipticity data set is split by galaxy type, a shape measurement would yield uncorrelated noise estimates for each galaxy type. Note that since $C_{t}$, $S_{t}$ and $N_{t}$ are made up from four $n_\mathrm{bin}\times n_\mathrm{bin}$-matrices, their total size is $2n_\mathrm{bin}\times 2n_\mathrm{bin}$. The indices $i$ and $j$ are the tomographic bin numbers, therefore, just as in tomographic lensing, increasing the number of bins increases the number of possible correlations, and thus leads to a gain in statistical information. It should be noted that we assume the same ellipticity dispersion for both galaxy types. This assumption, however, can be relaxed without altering the method.

Statistical inference from the covariance $C_t(\ell)$ would yield exactly the same statistical errors on cosmological parameters as a measurement that would not differentiate between galaxy types. The corresponding Fisher-matrix $F_{\mu\nu}$ would be obtained from
\begin{equation}
F_{\mu\nu} =
f_\mathrm{sky}\sum_\ell\frac{2\ell+1}{2}
\trace\left(C_t^{-1}(\ell)\partial_\mu S_t(\ell)\:C_t^{-1}(\ell)\partial_\nu S_t(\ell)\right),
\end{equation}
and the signal to noise-ratio of the non-randomness of the ellipticity field would be established at a significance $\Sigma$ of
\begin{equation}
\Sigma^2 =
f_\mathrm{sky}\sum_\ell\frac{2\ell+1}{2}
\trace\left(C_t^{-1}(\ell) S_t(\ell)\: C_t^{-1}(\ell) S_t(\ell)
\right).
\end{equation}

The two maps $e_{\mathrm{s},i}$ and $e_{\mathrm{e},i}$ can be linearly superposed in order to up- or downweight the relative contributions of the spectra $C^{\gamma}_{ij}(\ell)$, $C^{\mathrm{e},II}_{ij}(\ell)$, $C^{\mathrm{e},GI}_{ij}(\ell)$ and $C^{\mathrm{s},II}_{ij}(\ell)$: We incorporate this by rotation with a mixing angle $\alpha$,
\begin{align}
e_{+,i} & = +\cos\alpha\:e_{\mathrm{s},i} + \sin\alpha\:e_{\mathrm{e},i},\\
e_{-,i} & = -\sin\alpha\:e_{\mathrm{s},i} + \cos\alpha\:e_{\mathrm{e},i}\; .
\label{eq_linear_combination}
\end{align}
This change of basis amounts to an orthogonal transformation of the data, $C_{ij}^{\pm}(\ell) = U C_t(\ell) U^T$, with the orthogonal matrix $U^{-1} = U^T$ given by eq. (\ref{eq_linear_combination}). Since the statistical errors encoded in $F_{\mu\nu}$ and the signal to noise-ratio $\Sigma$ are both trace-relationships which are invariant under orthogonal transformations, it would not have any influence on inference from a weighted measurement either. In other words, the mapping is one-to-one and thus preserves all the information content by definition.

But if one now focuses on the statistical property of a single field, for instance $\epsilon_{+,i}$, it is possible to influence the relative contribution of lensing and the intrinsic alignment terms by a suitable choice of $\alpha$: Of course this can be trivially achieved by setting $\alpha=0$, in which case there are only spiral galaxies in the measurement whose shape correlations would be described by $C^\gamma_{ij}(\ell) + C^{\mathrm{s},II}_{ij}(\ell)$, or by choosing $\alpha=\pi/2$, which eliminates all galaxies apart from the elliptical ones with the the shape correlation $C^\gamma_{ij}(\ell) + 2C^{\mathrm{e},GI}_{ij}(\ell) + C^{\mathrm{e},II}_{ij}(\ell)$. 

Interestingly, there is a choice for $\alpha$ that corresponds to the case where correlations involving weak lensing $C^\gamma_{ij}(\ell)$ and $C^{\mathrm{e},GI}_{ij}(\ell)$ are completely removed from the data: In fact, the correlation $C^{++}_{ij}(\ell)$ is given by
\begin{equation}\label{eq:Cpp}
\begin{split}
C^{++}_{ij}(\ell)& =
(n_s \cos \alpha + n_e \sin \alpha )^2\: C^\gamma_{ij}(\ell)  \\
& + 2(n_sn_e\cos\alpha\sin\alpha + n_e^2\sin^2\alpha)\: C^{\mathrm{e},GI}_{ij}(\ell) \\ 
& +
n_s^2\cos^2\alpha\: C^{\mathrm{s},II}_{ij}(\ell) +
n_e^2\sin^2\alpha\: C^{\mathrm{e},II}_{ij}(\ell)\; .
\end{split}
\end{equation}
The prefactors of the lensing signal can now be set to zero with a suitable choice of $\alpha$, eliminating weak lensing from the measurement. Incidentally, the prefactor $\cos\alpha\sin\alpha+\sin^2\alpha$ will then be zero as well, which cancels the cross-correlation between lensing and the intrinsic shapes of elliptical galaxies and keeps only intrinsic alignments. For the case $n_s = n_e$ one would select $\alpha$ to be $3\pi/4$, and the two alignment contributions $C^{\mathrm{s},II}_{ij}(\ell)$ and $C^{\mathrm{e},II}_{ij}(\ell)$ enter $C^{++}_{ij}(\ell)$ with the same weight, so one effectively retains only $C^{\mathrm{s},II}_{ij}(\ell) + C^{\mathrm{e},II}_{ij}(\ell)$ from $C^{++}_{ij}(\ell)$ with the choice $\alpha=3\pi/4$. Note that even if $n_s\neq n_e$, which will be generally the case, one will find an $\alpha$ such that all lensing contributions vanish.
On the level of the ellipticity maps, the value $\alpha=3\pi/4$ would set in the linear combination~(\ref{eq_linear_combination}) $\cos\alpha=-1/\sqrt{2}$ and $\sin\alpha=+1/\sqrt{2}$, which cancels $\gamma$ from $e_{+,i}$, $e_{+,i} = (e_{\mathrm{s},i} - e_{\mathrm{e},i})/\sqrt{2} = (\epsilon_{\mathrm{s},i} - \epsilon_{\mathrm{e},i})/\sqrt{2}$.
It should be noted that eq. (\ref{eq:Cpp}) assumes a particular alignment model, especially that there will be no intrinsic ellipticity correlations between spiral and elliptical galaxies and that the $GI$ term for spiral galaxies vanishes. However, even if we consider an alignment model which allows all possible correlations, the split can still be performed. In particular, an intrinsic alignment cross-correlation between spiral and elliptical galaxies, $C^{\mathrm{se},II}_{ij}$, would appear in eq. \ref{eq:Cpp} with a prefactor of $(2 n_s n_e \sin \alpha \cos \alpha)$, whilst a $GI$ term for the spiral galaxies $C^{\mathrm{s},GI}_{ij}$ would be weighted with $2(n_s^2 \cos^2\alpha + n_s n_e \sin \alpha \cos\alpha)$ in turn.

Peculiarly, these factors vanish always at the same choice of $\alpha$ that also eliminates $C^{\mathrm{s},II}_{ij}$, e.g. $\alpha = \pi/2$ for $q = 0.5$. In general, thus, it is always possible to remove all spiral alignment contributions at once using this method -- at the price of statistically diminishing the lensing signal and boosting the relative contribution of elliptical galaxies' alignment.

The aforementioned point that eliminates lensing completely -- meaning choosing $\alpha$ to be $3\pi/4$ for $q = 0.5$ -- still holds as well: The spiral $GI$ contributions vanish here alongside the elliptical $GI$ signal, whilst a spiral-elliptical cross-alignment would remain, albeit with a negative sign.

\begin{figure}
\begin{center}
\includegraphics[width = \columnwidth]{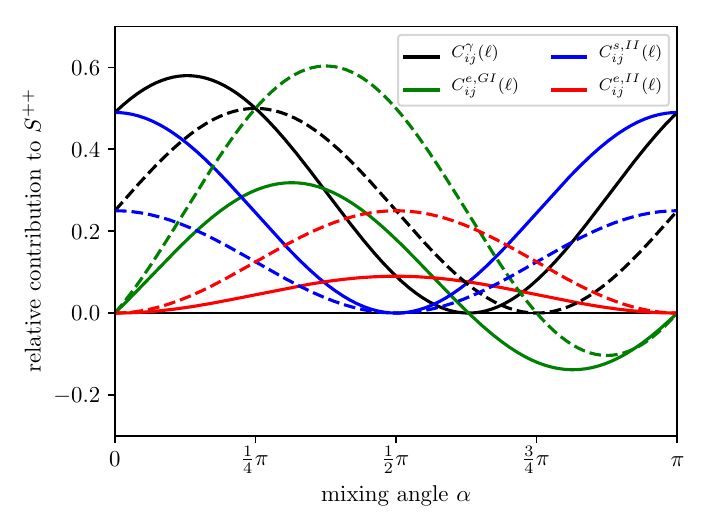}
\caption{Relative contributions of weak lensing and intrinsic alignments to the spectrum $C^{++}_{ij}(\ell)$ as a function of mixing angle $\alpha$: weak gravitational lensing $C^\gamma_{ij}(\ell)$ (black), $C^{\mathrm{e},GI}_{ij}(\ell)$ (green), $C^{\mathrm{s},II}_{ij}(\ell)$ (red line) and $C^{\mathrm{e},II}_{ij}(\ell)$ (blue). Dashed lines correspond to a $q = 0.5$, while solid lines have $q=0.7$. At $\alpha=3\pi/4$, the contributions involving weak lensing vanish for $n_s=n_e$.}
\label{fig_weighting}
\end{center}
\end{figure}

As shown by \autoref{fig_weighting}, the choice of $\alpha=\pi/4$ would boost the lensing signal relative to the intrinsic alignments for the case $n_s = n_e$, but would acquire nonzero contributions from every other correlation, with the peculiarity that the relative weighting of the two intrinsic alignments are identical. On the level of the ellipticity maps, $\alpha=\pi/4$ sets $\cos\alpha = \sin\alpha=1/\sqrt{2}$ such that $\mathrm{e}_{+,i} = (2\gamma_i + \epsilon_{\mathrm{s},i} + \epsilon_{\mathrm{e},i})/\sqrt{2}$, showing the maximal enhancement of $\gamma$, but with a non-zero contribution from all other ellipticity fields. We will investigate if other choices of $\alpha$ will lead to a better relative suppression of intrinsic alignments in the lensing signal, but one should point out that this procedure is not general and will depend on the particular lensing and intrinsic alignment spectra. For completeness, evaluating the remaining correlations yields
\begin{equation}
\begin{split}
C^{--}_{ij}(\ell) & =
(n_s \sin \alpha - n_e \cos \alpha)^2\: C^\gamma_{ij}(\ell) \\ &
+2(n_e^2\cos^2\alpha - n_sn_e\cos\alpha\sin\alpha) C^{\mathrm{e},GI}_{ij}(\ell)\: \\ & +
n_s^2\sin^2\alpha\: C^{\mathrm{s},II}_{ij}(\ell) +
n_e^2\cos^2\alpha\: C^{\mathrm{e},II}_{ij}(\ell)\; ,
\end{split}
\end{equation}
and
\begin{equation}
\begin{split}
C^{+-}_{ij}(\ell) & =
\left(n_sn_e(\cos^2\alpha -\sin^2\alpha) + \cos\alpha\sin\alpha(n_e^2-n_s^2)\right)\: C^\gamma_{ij}(\ell) \\
& +
\left(n_sn_e(\cos^2\alpha-\sin^2\alpha) + 2n_e^2\cos\alpha\sin\alpha\right)\:C^{\mathrm{e},GI}_{ij}(\ell) \\
& +\cos\alpha\sin\alpha\: \left(n_e^2C^{\mathrm{e},II}_{ij}(\ell) - n_s^2C^{\mathrm{s},II}_{ij}(\ell)\right)\; ,
\end{split}
\end{equation}
showing that there identical optimised choices for $\alpha$ from $C^{--}_{ij}(\ell)$. Likewise, the contribution $N_t(\ell)$ to the covariance matrix due to shape noise changes under rotations by $\alpha$, implying for the entries
\begin{equation}
N_{ij}^{++}(\ell) = \sigma_\epsilon^2\:n_\mathrm{bin}\:\left(n_s\cos^2\alpha + n_e\sin^2\alpha\right)\; ,
\end{equation}
as well as
\begin{equation}
N_{ij}^{--}(\ell) = \sigma_\epsilon^2\:n_\mathrm{bin}\:\left(n_s\sin^2\alpha + n_e\cos^2\alpha\right)\; ,
\end{equation}
and finally
\begin{equation}
N_{ij}^{+-}(\ell) = \sigma_\epsilon^2\:n_\mathrm{bin}\:\left(n_e - n_s\right)\cos\alpha\sin\alpha\; ,
\end{equation}
which effectively corresponds to the expressions for $C_{ij}^{++}(\ell)$, $C_{ij}^{--}(\ell)$ and $C_{ij}^{+-}(\ell)$ if one sets the lensing effect to zero and replaces the intrinsic alignment spectra with a Poissonian noise term. With the spectra reflecting the number of galaxy pairs and the shape measurement noise being proportional to the number of galaxies on retains a correct relative normalisation of all spectra and noise terms. In this work, we will be considering $q=0.7$, i.e. $70$ per cent of the total galaxy sample are spirals. 

\subsection{Misclassification}
\begin{figure}
\begin{center}
\includegraphics[width=\columnwidth]{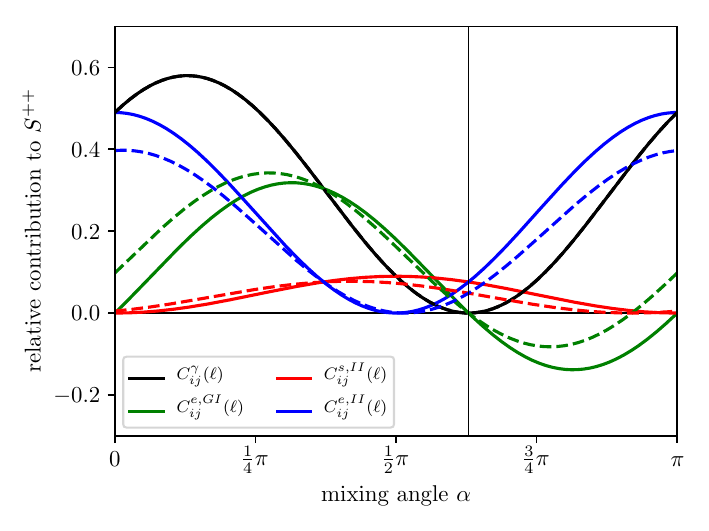}
\caption{Relative contributions of weak lensing and intrinsic alignments to the spectrum $C^{++}_{ij}(\ell)$ as a function of mixing angle $\alpha$. The colour scheme is the same as in \autoref{fig_weighting}. Here $q = 0.7$ for both dashed and solid curves. However, here, dashed curves have a misclassification rate $p(b|r) = p(r|b) = 0.1$, while the solid curves represent all galaxies identified correctly.}
\label{fig_weighting_missclassification}
\end{center}
\end{figure}

Imperfections in the classification of galaxies can be incorporated by introducing conditional probabilities of the type $p(b|r)$ indicating that an elliptical galaxy (with the label $r$, due to the red colour) is wrongly classified as a spiral galaxy (labelled $b$, due to the blue colour). Consequently, the number of observed spiral galaxies $n_b$ and observed elliptical galaxies $n_r$ is given in terms of the true number of spiral galaxies $n_s$ and elliptical galaxies $n_e$ by
\begin{align}
n_b & = p(b|b) n_s + p(b|r) n_e,\\
n_r & = p(r|b) n_s + p(r|r) n_e,
\end{align}
where the normalisation conditions $p(b|r) + p(r|r) = 1$ and $p(r|b) + p(b|b) = 1$ make sure that the total number of galaxies is conserved, $n = n_b+n_r = n_s + n_e$. A misclassification would take place if $p(b|r)$ and $p(r|b)$ were not equal to zero, so in fact only two of the conditional probabilities are independent variables. As a consequence of the misclassification one would obtain $(i)$ wrong amplitudes for the ellipticity correlations of a given galaxy type, $(ii)$ additional contributions from the respective other galaxy type with possible cross-correlations with the lensing signal, and $(iii)$ one would estimate these ellipticity spectra on the basis of the wrongly inferred number of galaxy pairs. The observed ellipticity fields in the tomographic bin $i$ now contain contributions from wrongly classified galaxies,
\begin{align}
e_{b,i} & = \gamma_i + p(b|b) \epsilon_{s,i} + p(b|r) \epsilon_{e,i},\\
e_{r,i} & = \gamma_i + p(r|b) \epsilon_{s,i} + p(r|r) \epsilon_{e,i}.
\end{align}
Consequently, the covariance matrix $C_f(\ell) = S_f(\ell) + N_f(\ell)$ reads
\begin{equation}
C_f(\ell) =
\left(
\begin{array}{ll}
C^{bb}_{ij}(\ell) & C^{br}_{i j^\prime}(\ell)\\
C^{rb}_{i^\prime j}(\ell) & C^{rr}_{i^\prime j^\prime}(\ell)
\end{array}
\right),
\end{equation}
whose signal part $S_f(\ell)$ contains contributions from wrongly classified galaxies. The components of matrix in the colour basis assume the following form
\begin{equation}
\begin{split}
S^{bb}_f(\ell) & = n_b^2\big[C^\gamma_{ij}(\ell) + 2p(b|r)C^{e,GI}_{ij}(\ell) \\ & + p(b|b)^2C^{s,II}_{ij}(\ell) + p(b|r)^2C^{e,II}_{ij}(\ell)\big] \\
S^{br}_f(\ell) & = n_bn_r\big[C^\gamma_{ij^\prime}(\ell) + C^{e,GI}_{ij^\prime}(\ell) \\
& + p(b|b)p(r|b)C^{s,II}_{ij^\prime}(\ell) + p(b|r)p(r|r)C^{e,II}_{ij^\prime}(\ell)\big] \\
S^{rb}_f(\ell) & = n_bn_r\big[C^\gamma_{i^\prime j}(\ell) + C^{e,GI}_{i^\prime j}(\ell) \\
& + p(b|b)p(r|b)C^{s,II}_{i^\prime j}(\ell) + p(b|r)p(r|r)C^{e,II}_{i^\prime j}(\ell)\big] \\
S^{rr}_f(\ell) & = n_r^2\big[C^\gamma_{i^\prime j^\prime}(\ell) + 2p(r|r)C^{e,GI}_{i^\prime j^\prime}(\ell) \\
& + p(r|b)^2C^{s,II}_{i^\prime j^\prime}(\ell) + p(r|r)^2C^{e,II}_{i^\prime j^\prime}(\ell)\big]
\end{split}
\label{eqn_sf}
\end{equation}
and whose noise part $N_f(\ell)$ needs to be updated because the numbers of blue and red galaxies are not equal to the numbers of spirals and ellipticals if $p(r|b)\neq 0$ or $p(b|r)\neq 0$,
\begin{equation}
N_f(\ell) = \sigma_\epsilon^2\:n_\mathrm{bin}\:
\left(
\begin{array}{ll}
n_b\delta_{ij} & 
0 \\
0 & 
n_r\delta_{i^\prime j^\prime}
\end{array}
\right).
\label{eqn_nf}
\end{equation}
It should be noted that equations~(\ref{eqn_sf}) and~(\ref{eqn_nf}) consistently reduce to equations~(\ref{eqn_st}) and~(\ref{eqn_nt}) if there is no misclassification, $p(r|b) = p(b|r) = 0$, such that the covariances are identical, $C_t(\ell) = C_f(\ell)$. Finally one would again rotate $C_f(\ell)$ into the new basis with the orthogonal matrix $U$. We will use for illustration purposes misidentification rates of $p(b|r) = p(r|b) = 0.1$. In \autoref{fig_weighting_missclassification} we show the influence of the misclassification: the dashed line shows a case with all galaxies being identified correctly, while the solid curves have misclassification of 10 $\%$. Clearly the lensing signal itself is unaffected as it does not depend on the colour of the galaxy. The other contributions however change in amplitude and dependence on $\alpha$. It is worth noting that the position where lensing and the $GI$ term vanish does not depend on the misclassification rate but only on the ratio of elliptical and spiral galaxies in the survey.

\subsection{Separating out intrinsic alignments}
The signal to noise-ratio $\Sigma$ of the isolated intrinsic alignments $C^{s,II}_{ij}(\ell) + C^{e,II}_{ij}(\ell)$ can be determined to be $\Sigma\simeq 60$ for Euclid: Focusing on the contribution $e_{+,i}$ after setting $\alpha=3\pi/4$ for the case $n_s=n_e$ defines the signal covariance $S^{++}_{ij}(\ell)$ as the upper left $n_\mathrm{bin}\times n_\mathrm{bin}$-block of $S_t(\ell)$. Consequently, the significance for measuring the intrinsic alignment contribution $C^{s,II}_{ij}(\ell) + C^{e,II}_{ij}(\ell)$ is given by
\begin{equation}
\Sigma^2 = f_\mathrm{sky}\sum_\ell\frac{2\ell+1}{2}
\trace\left(C_{++}^{-1}(\ell)S^{++}(\ell)\:C_{++}^{-1}(\ell)S^{++}(\ell)\right).
\label{eqn_sigma_ia}
\end{equation}
By isolating the $C_{++}$ components we throw away the information contained in the other components of the covariance.

\begin{figure}
\begin{center}
\includegraphics[width=\columnwidth]{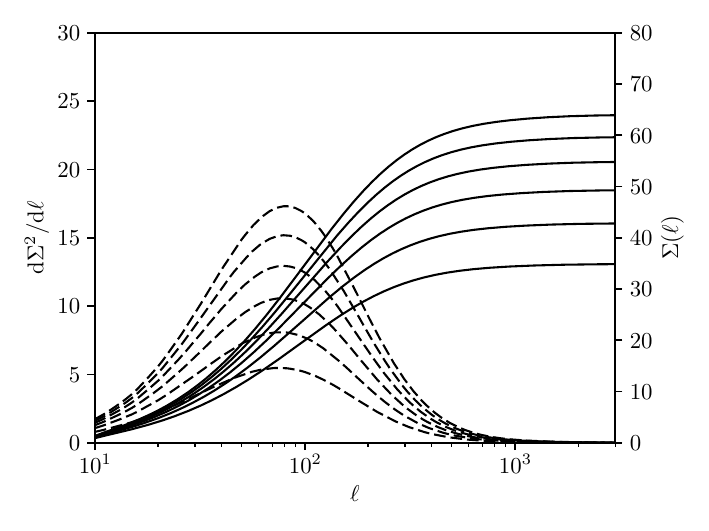}
\caption{Cumulative $\Sigma(\ell)$ (solid lines) and differential signal to noise-ratio $\dd\Sigma/\dd\ell$ (dashed lines) for the measurement of the intrinsic alignment spectra $C^{\mathrm{s},II}_{ij}(\ell) + C^{\mathrm{e},II}_{ij}(\ell)$ as defined by equation~\ref{eqn_sigma_ia} for a tomographic ellipticity measurement for $n_\mathrm{bin} = 2\ldots7$ bins for a survey like Euclid's.}
\label{fig_clean_ia}
\end{center}
\end{figure}

\autoref{fig_clean_ia} shows in detail the attainable significance $\Sigma$ for the intrinsic alignment contribution, both cumulatively and differentially, as a function of multipole moment $\ell$ and for different number of tomographic bins. Clearly, intrinsic alignments generate a significant signal in Euclid's survey and the magnitude of the alignments, both for spiral and elliptical galaxies, allows a determination of parameters of the alignment model at percent precision. It is also interesting to see that most of the signal is picked up at relatively low multipoles, showing that the assumption of Gaussian random fields can indeed be used. The reason for this is that the intrinsic alignment signal only probes the auto-correlation of different bins and thus the shape-noise dominates the high multipole moments, which becomes dominant earlier compared to the $II$ alignments due to their lower amplitude. Tomography is a large factor in attaining a high signal to noise ratio and therefore, in investigating intrinsic alignment models, as $\Sigma$ increases from 35 with 2 bins to over 60 with 7 bins. These numbers are in agreement with forecasts on alignment amplitudes: As the alignment signal is proportional to $A$ or $D$, the signal to noise-ratio corresponds to the inverse statistical error on these prefactors, such that one should achieve \%-level precision from a measurement yielding $100\sigma$ of statistical significance. It is remarkable that this level can be reached even if one considers a combination of cosmological probe and a complex cosmological model \citep{kitching_3d_2014, merkel_parameter_2017}.
\autoref{fig_clean_ia} shows in detail the attainable significance $\Sigma$ for the intrinsic alignment contribution, both cumulatively and differentially, as a function of multipole moment $\ell$ and for different number of tomographic bins. Clearly, intrinsic alignments generate a significant signal in Euclid's survey and the magnitude of the alignments, both for spiral and elliptical galaxies, allows a determination of parameters of the alignment model at percent precision. It is also interesting to see that most of the signal is picked up at relatively low multipoles, showing that the assumption of Gaussian random fields can indeed be used. The reason for this is that the $II$-part of intrinsic alignment signal only probes the auto-correlation of different bins (since IA is a small scale effect) and thus the shape-noise dominates the high multipole moments, which becomes dominant earlier compared to the $II$ alignments due to their lower amplitude. Tomography is a large factor in attaining a high signal to noise ratio and therefore, in investigating intrinsic alignment models, as $\Sigma$ increases from 35 with 2 bins to over 60 with 7 bins. These numbers are in agreement with forecasts for Euclid-like surveys on alignment amplitudes \citep[e.g.][]{tugendhat_angular_2018}: As the alignment signal is proportional to $A$ or $D$, the signal to noise-ratio corresponds to the inverse statistical error on these prefactors, such that one should achieve \%-level precision from a measurement yielding $100\sigma$ of statistical significance. It is remarkable that this level can be reached even if one considers a combination of cosmological probe and a complex cosmological model \citep{kitching_3d_2014, merkel_parameter_2017}

We quantify the difference between the ideal spectra and the one containing wrongly classified galaxies with the average $\bra\Delta\chi^2\ket$ between the two models. For that purpose, we identify the $C^{++}_{ij}(\ell)$ contribution from the rotated covariance matrix $C_t(\ell)$ and the corresponding $X^{++}_{ij}(\ell)$ from the rotated covariance matrix $C_f(\ell)$ including shape correlations of wrongly identified galaxies, by setting $\alpha = 3\pi/4$. 

The $\bra\chi^2_t\ket$-value of the true model is on average given by
\begin{equation}
\bra\chi^2_t\ket = f_\mathrm{sky}\sum_\ell(2\ell+1)\:\trace\left(\ln C^{++}(\ell) + \mathrm{id}\right).
\end{equation}
Here, $\mathrm{id}$ stands for the identity matrix. The average $\bra\chi^2_f\ket$ of the wrongly classified model can be computed with
\begin{equation}
\bra\chi^2_f\ket = f_\mathrm{sky}\sum_\ell(2\ell+1)\:\trace\left(\ln X^{++}(\ell) + X_{++}^{-1}(\ell)C^{++}(\ell)\right),
\end{equation}
such that the difference $\bra\Delta\chi^2\ket = \bra\chi^2_f-\chi^2_t\ket$ between the true and false model in the light of the on average expected data $C_t$ yields
\begin{equation}
\bra\Delta\chi^2\ket =
f_\mathrm{sky}\sum_\ell(2\ell+1)\:
\left(\ln\left(\frac{\mathrm{det}\:X^{++}(\ell)}{\mathrm{det}\:C^{++}(\ell)}\right) +
\trace\left(X_{++}^{-1}(\ell)C^{++}(\ell)\right)-n_\mathrm{bin}\right),
\end{equation}
where we again assume statistical homogeneity of the $e_{+,i}$ and $e_{-,i}$-fields, and a number of $2\ell+1$ statistically uncorrelated modes on a given angular scale $\ell$ as a consequence of statistical isotropy. We scale the resulting $\chi^2$-values or signal to noise-ratios $\Sigma$ by $\sqrt{f_\mathrm{sky}}$ if the sky coverage is incomplete. If $X^{++}(\ell)=C^{++}(\ell)$, then $\bra\Delta\chi^2\ket=0$ because $\trace\left(X_{++}^{-1}C^{++}\right)=n_\mathrm{bin}$. In the relations above, $\mathrm{id}$ refers to the unit matrix in $n_\mathrm{bin}$ dimensions.

\begin{figure}
\begin{center}
\includegraphics[width=\columnwidth]{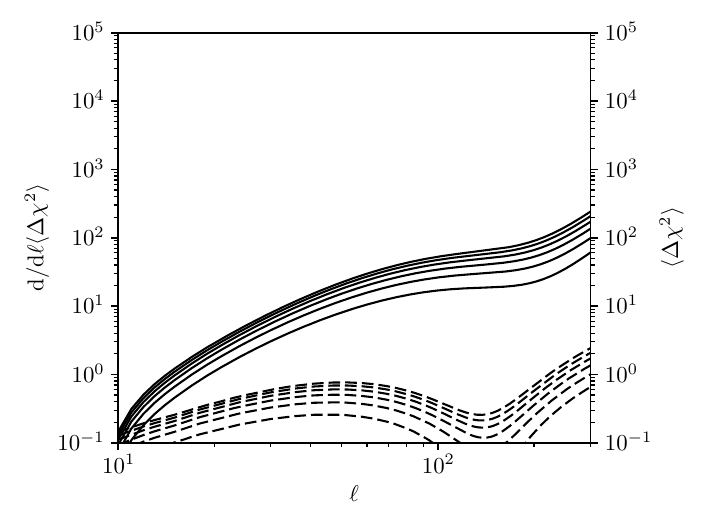}
\caption{Difference $\bra\Delta\chi^2\ket$ between the covariances $C^{++}(\ell)$ (where all galaxies are correctly identified) and $X^{++}(\ell)$ (with assumed misidentification rates of $p(b|r) = p(r|b) = 0.1$) as a function of multipole $\ell$ for $n_\mathrm{bin}=2\ldots7$ tomographic spectra.}
\label{fig_delta_chi2}
\end{center}
\end{figure}

First of all we expect the obtainable $\Sigma^2(\ell)$ to be smaller in the case where galaxies have been identified wrongly, since the relative contribution is smaller as it can be seen in \autoref{fig_weighting_missclassification}. This will lead to a difference  $\bra\Delta\chi^2\ket$ between the true model and the model containing wrongly identified galaxies by expressing the difference in the spectra in units of the cosmic variance, which we show in \autoref{fig_delta_chi2}. At $\ell \gsim 200$, the difference in $\chi^2$ continues to grow steadily due to the difference in the noise contributions, as the noise is dependent on the galaxy counts, which naturally differ if there is a misclassification of galaxy types. Furthermore, the loss of Gaussianity in the correlations at larger $\ell$ will lead to complicated cross--correlations between the alignment effects that have not been included in the calculation of $\bra\Delta \chi^2\ket$ or indeed in the considerations for a misclassified signal $S_f(\ell)$ (cf. \autoref{eqn_sf}).

\subsection{Boosting the weak lensing spectrum}
The weak lensing contribution $C^{\gamma}_{ij}(\ell)$ can be boosted relative to the intrinsic alignments $C^{s,II}_{ij}(\ell)$ and $C^{e,II}_{ij}(\ell)$ by choosing a value of $\alpha$ in the vicinity of $\pi/4$. Unlike the previous case, there is no complete cancellation of the intrinsic alignment contribution and one can only hope to optimise the measurement. This optimisation, however, does depend on the particular values of the lensing and alignment spectra and is therefore not model-independent. Because this is inevitably the case, there are two competing effects to be taken care of: Firstly, the amplitude of the contribution of the weak lensing spectrum to $C^{++}_{ij}(\ell,\alpha)$ changes as a function of $\alpha$, giving rise to a change in the statistical precision of the measurement encoded in the Fisher-matrix,
\begin{equation}
F_{\mu\nu}(\alpha) = f_\mathrm{sky}\sum_\ell\frac{2\ell+1}{2}
\trace\left(C^{-1}_{++}(\ell,\alpha)\partial_\mu S^{++}(\ell)\:
C^{-1}_{++}(\ell,\alpha)\partial_\nu S^{++}(\ell)\right),
\end{equation}
and resulting in changes in the marginalised statistical errors $\sigma_\mu^2(\alpha) = \left(F^{-1}(\alpha)\right)_{\mu\mu}$. We approximate derivatives $\partial_\mu S^{++}_{ij}(\ell)$ with the derivatives at $\alpha=\pi/4$ because weak lensing dominates the spectrum and the derivatives with respect to the cosmological parameters are identical for every choice of $\alpha$ in this limit.

Secondly, the relative contribution of the intrinsic alignment spectra change as well with $\alpha$ such that there is a varying amount of contamination of intrinsic alignments and a corresponding systematic error $\delta_\mu(\alpha)$. This systematic error is computed using the formalism of \citet{schafer_weak_2012} as an extension to \citet{taburet_biases_2009} and \citep{amara_systematic_2008} from 
\begin{equation}
\delta_\mu(\alpha) = \sum_\nu\left(G^{-1}(\alpha)\right)_{\mu\nu}a_{\nu}(\alpha)
\end{equation} 
with the matrix $G_{\mu\nu}$
\begin{equation}
\begin{split}
& G_{\mu\nu}(\alpha) = \\
 & = f_\mathrm{sky}\sum_\ell\frac{2\ell+1}{2}
\trace
\left[C_{++}^{-1}(\alpha)\:\partial^2_{\mu\nu}C^{++}(\ell)\:\left(C_{++}^{-1}(\alpha)C^{++}(\ell)-\mathrm{id}\right)\right] \\
&- 
\trace
\left[C^{-1}_{++}(\alpha)\partial_\mu C^{++}(\ell)\:C^{-1}_{++}(\alpha)\partial_\nu C^{++}(\ell)\:\left(2C_{++}^{-1}(\alpha)C^{++}(\ell)-\mathrm{id}\right)\right],
\end{split}
\end{equation}
and the vector $a_\mu$
\begin{equation}
a_{\mu}(\alpha) = f_\mathrm{sky}\sum_\ell\frac{2\ell+1}{2}
\trace\left[C^{-1}_{++}(\alpha)\partial_\mu C^{++}(\ell)\left(\mathrm{id}-C_{++}^{-1}(\alpha)C^{++}(\ell)\right)\right]
\end{equation}
which are both functions of the mixing angle $\alpha$. We collect multipoles up to $\ell_\mathrm{max} =2500$.

\begin{figure}
\begin{center}
\includegraphics[width=\columnwidth]{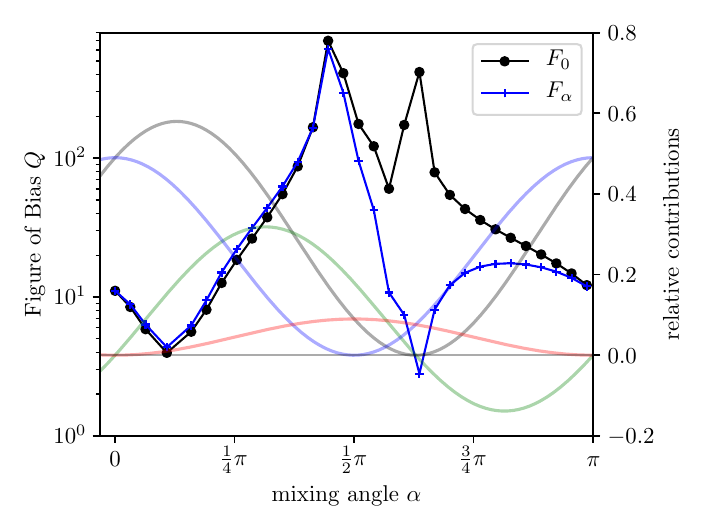}
\caption{Total bias $Q(\alpha)$ in units of the statistical error, $Q^2=\sum_{\mu\nu}F_{\mu\nu}\delta_\mu\delta_\nu$ as a function of the mixing angle $\alpha$, where the minimum indicates the value of $\alpha$ that is able to yield the smallest possible systematic error corresponding to the cleanest weak lensing spectrum. $F_{0}$ (black) corresponds to a derivative at $\alpha=\pi/4$, whereas $F_{\alpha}$ in blue shows the same with derivative taken at the corresponding $\alpha$. The translucent lines in the background are taken from \autoref{fig_weighting} as a guideline where IAs vanish and where they dominate.}
\label{fig_biases}
\end{center}
\end{figure}

A convenient way for expressing the magnitude of the systematic error in units of the statistical error is the figure of bias $Q(\alpha)$,
\begin{equation}
Q^2(\alpha) = \sum_{\mu\nu}F_{\mu\nu}(\alpha)\:\delta_\mu(\alpha)\delta_\nu(\alpha),
\end{equation}
which is related to the Kullback-Leibler divergence $D_\mathrm{KL}$ for Gaussian likelihoods under the assumption of constant covariances,
\begin{equation}
D_\mathrm{KL} = \frac{Q^2}{2}. 
\end{equation}
\autoref{fig_biases} shows $Q$ as a function of $\alpha$: Clearly, there is an optimised choice of $\alpha$ that reduces the intrinsic alignment contribution in order to yield an unbiased measurement of the cosmological parameter set, which we demonstrate by computing $Q$ in scanning through all possible choices of $\alpha$. This optimal choice of $\alpha$ is where the curves dip between $0$ and $\pi/4$, where the bias is roughly halved. We calculate the figure of bias $Q$ both with the systematic errors taken in relation to the statistical errors at $\alpha = 0$, $F_0$, in black and with the Fisher matrix taken at the respective $\alpha$, $F_\alpha$, in blue. It becomes clear that as long as there is a significant lensing contribution, up until $\alpha \simeq \pi/2$, the differences are negligible. As soon as the contributions from intrinsic alignments become comparable to the one from weak lensing, the figure of bias forks: while comparing the systematic errors to a co--evolving statistical error, the figure of bias becomes increasingly small. This is not due to a diminishing bias but rather due to extremely large statistical errors as the information from lensing disappears. For a constant statistical error taken at $\alpha=0$, the curve is therefore more representative of the actual precision of the measurement. As $\alpha$ approaches $\pi$, the curves start to merge again.

Typically, relative contributions of order ten percent of the intrinsic alignment signal to weak lensing cause figures of bias $Q$ of the order up to a few hundred, which can be controlled by our technique. It would be interesting to propagate intrinsic alignments with other systematic effects through the parameter estimation process as outlined by \citet{cardone_power_2014}.

\begin{figure}
\begin{center}
\includegraphics[width=\columnwidth]{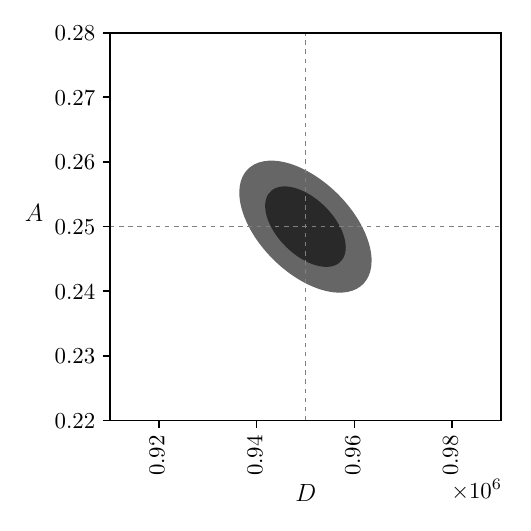}
\caption{Fisher-constraints of the model parameters $A$ and $D$ for the quadratic and linear alignment models respectively for $n_\mathrm{bin}=7$ in Euclid.}
\label{fig_fisherAD}
\end{center}
\end{figure}

Lastly, we aim to constrain the parameters of the alignment models for the two types of galaxies, specifically the alignment amplitude $D$ that describes the elasticity of elliptical galaxies and the amplitude $A$ which describes the magnitude $A$ of the misalignment between tidal shear and inertia which is responsible for angular momentum generation in spiral galaxies. If one suppresses lensing, including $GI$-alignment, \autoref{fig_fisherAD} suggests that both alignment parameters can be constrained at the percent level with the Euclid data set without having to worry about biases due to gravitational lensing. These numbers result from a Fisher-matrix analysis for the parameters $A$ and $D$ with a choice of $\alpha$ that eliminates lensing, and fitting the corresponding intrinsic ellipticity spectra to the remainder, while taking account of cosmic variance and shape noise contributions and keeping all cosmological parameters fixed to their fiducial values.

\section{Summary}\label{sect_summary}
Subject of this investigation was the use of colour information to differentiate between intrinsic ellipticity correlations of galaxies and the weak gravitational lensing signal. We operate under the assumptions that $(i)$ the large-scale structure follows Gaussian statistics on large scales, $(ii)$ intrinsic shapes of elliptical galaxies follows the linear tidal shearing model, $(iii)$ intrinsic shapes of spiral galaxies follows the quadratic tidal torquing model, $(iv)$ gravitational lensing is universal and linear in the tidal shear, and $(v)$ it is possible to classify galaxies on the basis of colour information to obey one of the two alignment mechanisms. $(vi)$ shape and redshift measurements for the cosmic shear analysis are under control.
Based on the fact that lensing is universal and affects all galaxy shapes in an identical way irrespective of galaxy type it is possible to find linear combinations of ellipticity fields measured for spiral and elliptical galaxies that do not contain any lensing signal and only retain shape correlations due to intrinsic alignments. It should be emphasised that this is possible without any assumption about the details of gravitational lensing or of the intrinsic alignment models. Intrinsic alignments that have been separated in this way from the lensing signal, can be measured on the basis of Euclid's weak lensing data set with a high statistical significance.
The subject of this investigation was the use of colour information to differentiate between intrinsic ellipticity correlations of galaxies and the weak gravitational lensing signal. We operate under the assumptions that $(i)$ the large-scale structure follows Gaussian statistics on large scales, $(ii)$ intrinsic shapes of elliptical galaxies follows the linear tidal shearing model, $(iii)$ intrinsic shapes of spiral galaxies follows the quadratic tidal torquing model, $(iv)$ gravitational lensing is universal and linear in the tidal shear, and $(v)$ it is possible to classify galaxies on the basis of colour information to obey one of the two alignment mechanisms. Based on the fact that lensing is universal and affects all galaxy shapes in an identical way irrespective of galaxy type it is possible to find linear combinations of ellipticity fields measured for spiral and elliptical galaxies that do not contain any lensing signal and only retain shape correlations due to intrinsic alignments. It should be emphasised that this is possible without any assumption about the details of gravitational lensing or of the intrinsic alignment models. Intrinsic alignments that have been separated in this way from the lensing signal, can be measured on the basis of Euclid's weak lensing data set with a high statistical significance.

\begin{enumerate}
\item{As alignment models we consider tidal shearing for elliptical galaxies and tidal torquing for spiral galaxies, and compute the resulting tomographic angular ellipticity spectra including the non-zero cross-correlation between the shape of elliptical galaxies and weak lensing. Both models have a single free parameter each, which we determined from weak lensing data and from numerical simulations, respectively. In analysing data, we showed that Euclid will allow their measurement at a level of a few percent. In that, we assume constant and scale-independent alignment parameters. The forecasted statistical precision would allow the investigation of alignment models with Euclid's weak lensing data set, and shed light on $(i)$ the average misalignment of angular momenta with tidal shear fields and ($ii$) the reaction of a virialised structure to external tidal shear fields.}
\item{We develop a statistical method which allows the separation between weak lensing and both alignment types on the basis of colour or morphological information: We assume that elliptical galaxies, if they are correctly identified, obey exclusively linear tidal shearing as their alignment mechanism, while spiral galaxies are described by the quadratic tidal torquing model. Gravitational lensing is universal as it affects the shapes of spiral and elliptical galaxies identically.}
\item{All mitigation and suppression techniques have in common that statistical precision is traded for systematical accuracy, and our method is no exception: Starting from shape catalogues measured for different types of galaxies it is possible to find linear combinations of the shape measurements that contain no spiral alignment, no elliptical alignment or no gravitational lensing, including in this case no lensing-alignment cross correlation either. With that in mind, it is possible to find a linear combination that eliminates lensing from the data set and leaves only contributions to the shape correlations that differ between elliptical and spiral galaxies, i.e. intrinsic alignments. Taking these cleaned correlation functions, they allow a measurement of the intrinsic alignment signal without any model assumptions to $\sim60\sigma$ of statistical precision for a tomographic survey such as Euclid's. For other surveys, such as LSST, we expect a significantly lower contributions of IA, since they are a) deeper and b) less wide than Euclid. The effect from a) is that the survey gathers information from redshifts where IA is much lower, as structure formation is less developed at those stages as well as the lensing efficiency dominates even more strongly. From b) we expect a lower statistical significance, as there is more shot noise, which also washes out IA contributions. Therefore, only a survey that clearly makes out IA contamination to a significant amount is relevant for this method.}
\item{An independent verification of the validity of the separation of lensing and IA contributions could possibly be done by measuring $B$-mode spectra and comparing those to the predicted $B$-modes (if any) of the underlying intrinsic alignment models. This is not particular to our choice of models but can in principle be done for any assumption of IA models, especially to verify or improve the model parameter measurement.}
\item{Suppression of intrinsic alignment contributions for achieving a bias-free weak lensing measurement is possible, but not completely. With an optimised choice of the mixing angle $\alpha$ one can reduce the systematical bias in units of the statistical error by a large margin, and we show that biases are reduced to amount to typically a few $\sigma$ for the full $\Lambda$CDM parameter set. We quantify the magnitude of the systematic error in units of the statistical error by the figure of bias $Q^2 = \sum_{\mu\nu}F_{\mu\nu}\delta_\mu\delta_\nu$, which takes care of the orientation of the systematic error $\delta_\mu$ with respect to the statistical degeneracies encoded in the Fisher-matrix $F_{\mu\nu}$. Incidentially, $Q^2/2$ corresponds to the Kullback-Leibler-divergence between the biased and unbiased likelihood $\mathcal{L}$.}
\item{Misclassification, i.e. non-zero probabilities $p(r|b)$ and $p(b|r)$ do not affect our conclusions strongly even for very high probabilities of misclassification on the order of ten per cent. This generous limit for misclassifications is also chosen to show that additional IA mechanisms that can be treated as misclassifications would not change our results meaningfully. We quantify this by computing the difference between the resulting covariance in terms of the $\chi^2$-statistic, where instrumental noise and cosmic variance are present as noise sources. At accessible scales below a few hundred in $\ell$, i.e. before the instrumental noise dominates, the integrated $\Delta\chi^2$ is a few ten, which, given the sensitivity of a weak lensing survey with respect to parameters such as $\sigma_8$ or $\Omega_m$, would give rise to biases. We estimate that the misidentification probabilities would need to be controlled to well below ten per cent for the biases not to exceed $1\sigma$ in terms of the statistical error in this case.}
\end{enumerate}

Precursing to the work presented here, there are quite a few other mitigation techniques: \citet{catelan_intrinsic_2001} proposed tomographic methods to reduce the $II$-alignment signal, by avoiding spatially close galaxies. This exploits the large correlation length of the weak lensing signal, compared to the intrinsic alignment signal. The drawback of this method is the increased cosmic variance and shape noise and thus a loss of statistical power. This technique was used by several authors \citep[e.g.][]{heymans_weak_2003, heymans_weak_2004, king_suppressing_2002, heymans_weak_2004, Takada2004, joachimi_intrinsic_2013, heymans_cfhtlens_2013}.

Another method is to construct a different weighting of the cosmic shear signal, to reduce the contamination by $GI$-alignments. This nulling technique was discussed by \citet{huterer_nulling_2005, joachimi_controlling_2010}. It has been proposed as well to null and boost magnification \citep{heavens_cosmic_2011, Schneider2014}, which can also bias the parameter inference process. Alternatively, one can also use self-calibration techniques, making use of additional information and the cross-correlation of cosmic shear and galaxy clustering, as well as galaxy clustering auto-correlations, which has been used by \citet{bernstein_dark_2004, bernstein_comprehensive_2009, zhang_proposal_2010} or use the large difference in the amplitudes of $E$- and $B$-mode spectra of intrinsic alignments and weak lensing \citep{crittenden_discriminating_2002, schaefer_angular_2015}. Our method could as well be extended bispectra of the ellipticity field \citep{shi_controlling_2010, merkel_theoretical_2014, larsen_intrinsic_2016}, either in order to make the method less dependent on the assumption of Gaussianity of the tidal shear field or by relaxing on the relationship that the observable shape depends on the tidal shear field in a linear or quadratic way. Taking this idea further, it would be very interesting to see if other cross-correlation measurements, for instance correlations between the CMB-lensing field and galaxy shapes at higher redshifts than the ones considered here, would constraint alignment processes as well \citep{hall_intrinsic_2014}.

In summary, there is a wealth of information that helps to differentiate intrinsic alignments and weak lensing, even without sacrificing statistical precision for accuracy. While our investigation assumes that it is possible to assign an alignment model to a given galaxy on the basis of its colour (or other morphological information), inclusion of higher-order statistics, $E/B$-mode decomposition or redshift weighting scheme should enable a thorough understanding of shape correlations. We would consider the possibility of eliminating the lensing signal including the $GI$-terms from the ellipticity correlation through a suitable choice of $\alpha$ very interesting for the model-independent detection of intrinsic alignments.

\section*{Acknowledgements}
TMT acknowledges a doctoral fellowship from Astronomisches Rechen-Institut in Heidelberg, and RFR's work is supported by a stipend of Landesgraduiertenf{\"o}rderung Baden-W{\"u}rttemberg from the graduate college {\em Astrophysics of cosmological probes of gravity}. BMS would like to thank the University of Auckland, in particular S. Hotchkiss and R. Easthers for their hospitality during the early stages of this project, and A. Nusser and V. Desjacques for their hospitality at the Technion in Haifa during the final writeup. We would like to thank H. Hildebrandt for a suggestion for values of $p(b|r)$ and $p(r|b)$. We would furthermore like to thank the anonymous referee for their invaluable corrections, suggestions for improvement, and patience.

\bibliographystyle{mnras}
\bibliography{references_sep}

\bsp
\label{lastpage}
\end{document}